\documentclass{article}

\usepackage[preprint]{neurips_2026}

\usepackage[utf8]{inputenc} % allow utf-8 input
\usepackage[T1]{fontenc}    % use 8-bit T1 fonts
\usepackage{hyperref}       % hyperlinks
\usepackage{url}            % simple URL typesetting
\usepackage{booktabs}       % professional-quality tables
\usepackage{amsfonts}       % blackboard math symbols
\usepackage{nicefrac}       % compact symbols for 1/2, etc.
\usepackage{microtype}      % microtypography
\usepackage{xcolor}         % colors

\usepackage{microtype}
\usepackage{graphicx}
\usepackage{subcaption}
\usepackage{booktabs} % for professional tables
\usepackage{tcolorbox}
\usepackage{hyperref}
\usepackage{enumitem}
\usepackage[utf8]{inputenc} % allow utf-8 input
\usepackage[T1]{fontenc}    % use 8-bit T1 fonts
\usepackage{hyperref}       % hyperlinks
\usepackage{url}            % simple URL typesetting
\usepackage{booktabs}       % professional-quality tables
\usepackage{amsfonts}       % blackboard math symbols
\usepackage{microtype}      % microtypography
\usepackage{xcolor}         % colors
\usepackage{graphicx} % for including images
\usepackage{subcaption}
\usepackage{wrapfig}
\usepackage{multirow}
\usepackage{color}
\usepackage{colortbl}
\usepackage{algorithmic}
\usepackage{amsmath}
\usepackage{algorithm}   % 必装
\usepackage{amsfonts}    % 支持\mathbb{I}
\renewcommand{\algorithmicrequire}{ \textbf{Input:}} %Use Input in the format of Algorithm
\renewcommand{\algorithmicensure}{ \textbf{Output:}}
\usepackage{bm}
\newcommand{\ie}{\textit{i.e., }}
\newcommand{\eg}{\textit{e.g., }}

\definecolor{mygray}{rgb}{0.9, 0.9, 0.9}

\title{Joint Optimization of Multi-agent Memory System}

\author{
Wenyu~Mao\thanks{This work was completed during the internship at ByteDance Seed. Email: wenyumao2@gmail.com}\textsuperscript{1},\,
Haoyang Liu\textsuperscript{1}\,
Haosong Tan\textsuperscript{1},\,
Yaorui Shi\textsuperscript{1},\,
Jiancan Wu\textsuperscript{1,2},\,
An Zhang\textsuperscript{1},\,
Xiang~Wang\thanks{Corresponding author. Email: xiangwang@ustc.edu.cn.}\textsuperscript{1}\,\\
\textsuperscript{1}
University of Science and Technology of China\\
\textsuperscript{2}Institute of Dataspace, Hefei Comprehensive National Science Center\\
}

\begin{document}

\maketitle

\begin{abstract}
Memory systems are critical for LLMs, mitigating context window limitations and supporting long-horizon user--LLM interactions. 
Such systems typically comprise multiple agents responsible for memory construction and retrieval. 
Existing approaches often optimize each agent independently under a shared global objective (\eg downstream QA accuracy), treating other agents as a static environment. 
However, this design has two key limitations: (1) independent optimization ignores inter-agent dependencies and lacks agents' co-adaptation, and (2) relying solely on sparse global rewards provides limited guidance for optimizing specialized agents and causes ambiguous credit assignment. These may ultimately limit agents' effective collaboration in the memory system.
To address these limitations, we propose \textbf{CoMAM}, a joint optimization framework that promotes collaboration among agents via end-to-end reinforcement learning and an adaptive credit assignment mechanism. 
Specifically, we model the multi-agent pipeline as a Markov decision process (MDP) to expose inter-agent dependencies during end-to-end training. 
Agents are then jointly optimized using a combination of their local task reward and an adaptively weighted global reward, enabling agents to co-adapt while receiving targeted feedback for their respective roles. 
Experiments show that CoMAM consistently outperforms leading memory systems, validating the effectiveness of the joint optimization framework.
\end{abstract}

\section{Introduction}

Memory systems \cite{Mem0,memorybank,hu2025memory} have emerged as a pivotal component for large language models (LLMs), enabling them to preserve long-horizon conversational history beyond the context window and selectively incorporate it for generation. 
Such systems are typically implemented as multi-agent pipelines, where \textit{construction agents} summarize and store user history into multi-granular memories (\eg long/short or coarse/fine) \cite{li2025cam,mirix,memory-alpha}, and \textit{retrieval agents} select and integrate relevant memories for downstream queries via RAG \cite{RAG} or search-based reasoning \cite{memagent,search-r1}. These agents are often instantiated with heterogeneous configurations to handle tasks of varying complexity, enabling flexible modular design and specialization.
To enhance the performance of memory systems, existing approaches \cite{mem-r1,mem1,GAM,memory-alpha} often optimize each agent independently via reinforcement learning under a shared global objective (\eg, QA accuracy on downstream tasks). They typically update one agent at a time, treating the others as fixed components of the environment.

\begin{figure}[t]
  \includegraphics[width=\columnwidth]{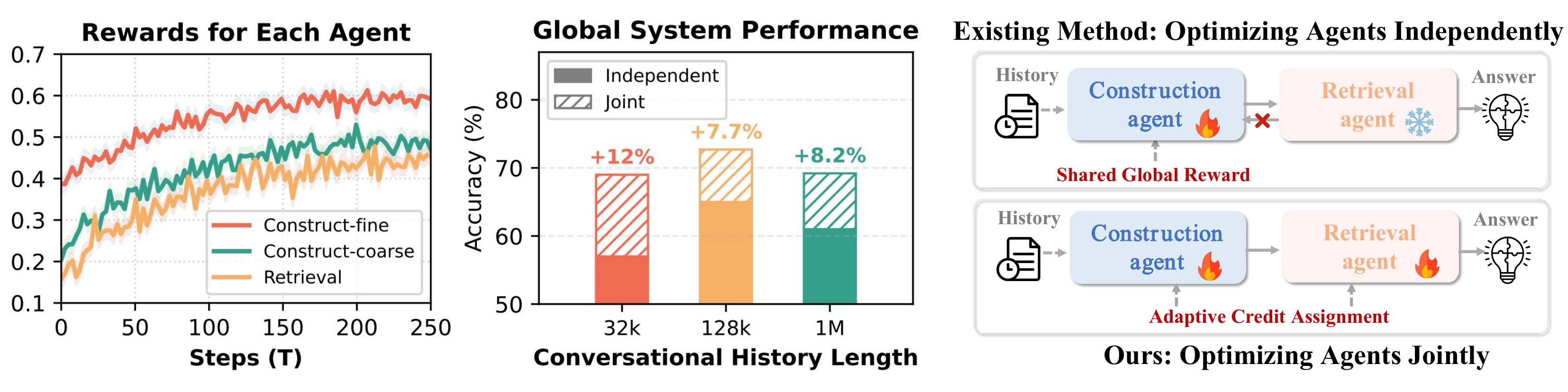}
  \caption{
  (\textbf{Left}) Training curves of coarse/fine memory construction and retrieval agents, which are optimized independently. 
  (\textbf{Middle}) Global system performance under independent and joint optimization across different context lengths. 
  (\textbf{Right}) Conceptual comparison between existing methods that optimize agents independently with shared global rewards and our joint optimization with adaptive credit assignment.
  }
  \label{fig:teaser_challenge}
  \vspace{-6mm}
\end{figure}

However, such independent optimization has two limitations and may lead to sub-optimal division of labor and collaboration \cite{local_optimalty,Optimas}. 
First, \textbf{ignorance of dependency among agents}: each agent is optimized while treating others as a static environment, preventing the adaptation to other agents’ evolving policies.
Second, \textbf{ambiguous credit assignment}: the shared global objective provides insufficient signal to attribute performance improvements to individual agents, which may hinder effective specialization.
As illustrated in Figure \ref{fig:teaser_challenge}, although agents can be well trained independently (rewards converged in the left figure), combining them directly leads to inferior system performance compared to joint optimization across different context lengths (lower QA accuracy in the middle figure). This gap highlights that independent optimization with the same global system objective is insufficient to achieve optimal system performance, motivating us to jointly optimize these heterogeneous agents in the memory system to facilitate collaboration.

To address these limitations, we propose \textbf{CoMAM}, a joint optimization framework that promotes collaboration among agents via end-to-end reinforcement learning, together with an adaptive credit assignment mechanism.
As illustrated in the right panel of Figure \ref{fig:teaser_challenge}, our approach differs from independent optimization by enabling simultaneous updates of heterogeneous agents with adaptively assigned rewards.
Specifically, to formulate a standard memory system for coarse/fine memory construction and memory retrieval, we design an extraction agent to construct fine-grained memories (\eg key events and facts), a profile agent to summarize coarse-grained memories (\eg users' stable preferences), and a retrieval agent to retrieve memories and answer queries. To model the inter-agent dependencies for co-adaption, we frame the heterogeneous multi-agent pipeline as a sequential Markov decision process (MDP), enabling agents to coordinate through their intermediate outputs. 
To alleviate ambiguous credit assignment during joint optimization, CoMAM assigns each agent a combination of its local task reward and an adaptively weighted global reward, providing more targeted feedback for different roles, driving collaboration for the global objective while maintaining task specialization.
Extensive experiments on long-horizon conversational tasks show CoMAM outperforms leading memory baselines, validating the efficacy of our joint optimization framework and adaptive credit assignment for multi-agent memory systems.

\section{Related work}
\label{sec:related_work}

\paragraph{Memory Systems for LLMs.}
To extend LLMs beyond their context window, recent work introduces memory systems to handle long-horizon user--LLM interactions \cite{hu2025memory,li2025cam,Mem0,mem-r1}. 
These systems are typically implemented as multi-agent pipelines \cite{GAM,li2025cam,memorybank,mem-r1}, where heterogeneous agents perform memory construction and retrieval, respectively.
For example, MIRIX \cite{mirix} employs a meta agent with multiple specialized agents for memory management. 
Mem$-\alpha$ \cite{memory-alpha} and Mem1 \cite{mem1} use RL to optimize memory construction and retrieval agents, respectively. Memory-R1 \cite{mem-r1} introduces two RL-trained agents for memory management and answer generation, which are optimized separately under a shared downstream objective.
% Despite these advances, most approaches optimize agents individually without explicitly considering inter-agent dependencies, especially in heterogeneous settings with diverse roles. In contrast, our work focuses on jointly optimizing multiple agents to enable coordinated co-evolution under adaptive credit assignment.

\paragraph{Optimization of Multi-agent Systems.}
Multi-agent systems (MAS) \cite{masrl_syrvey,self_play,Optimas} leverage multiple specialized agents with heterogeneous configurations to tackle complex tasks beyond the capacity of a single agent. 
Existing approaches typically optimize agents independently, either through prompt engineering \cite{Gorilla,avatar} or task-specific fine-tuning \cite{independent_mas,mem-r1}. 
Such independent optimization treats other agents as part of the environment and does not explicitly account for their inner dependence during training.
Consequently, recent research has shifted to RL-based joint training for end-to-end system optimization, such as MAPoRL \cite{MAPORL} and MARFT \cite{MARFT}.

\paragraph{Credit Assignment.}
Credit assignment \cite{credit_assign3,credit_assign5,credit_assign4} concerns properly attributing a team's global outcome to the specific agents in Multi-Agent Reinforcement Learning. This is challenging due to the interdependence among agents and environmental noise, which obscures each agent’s contribution.
A common strategy is to distribute global rewards uniformly across agents \cite{naive_assign,hong2025multi}. 
While simple, this leads to significant credit assignment ambiguity: agents cannot distinguish their specific contributions \cite{COMA,credit_assignment3}, resulting in high variance in policy updates and unstable convergence.

\section{Preliminary}
\subsection{Task Formulation for Multi-agent Memory System}
\label{sec:task_formulation}
We define the long-horizon user-LLM conversational history as $\mathcal{H} = \{h_1, \dots, h_N\}$, which consists of $N$-turn interactions that exceed the LLM's context window. The Memory System transforms the history $\mathcal{H}$ into multi-granular memories $\mathcal{M}$ that can be retrieved to answer users' subsequent queries $q$, realized by heterogeneous agents on construction and retrieval tasks \cite{hu2025memory}:
\begin{itemize}[leftmargin=*]
\item \textbf{Memory Construction:} The agents \cite{li2025cam} under policies $\{\pi_\theta^{\text{cons}_g}\}$, distill the raw conversational history $\mathcal{H}$ at different granularities $g$, finally constructing a multi-granular memory set $M$:
\begin{gather}
\mathcal{M}_g = \pi_\theta^{\text{cons}_g}(\mathcal{H}),\quad
\mathcal{M} = \{\mathcal{M}_g\}.
\end{gather}

\item \textbf{Memory Retrieval:} Given a subsequent query $q$ of downstream task, retrieval agents \cite{mem1} under policies $\{\pi_\theta^{\text{ret}}\}$ identify the $K$ most relevant memories from $\mathcal{M}$. Then, the agent with policy $\pi_\theta^{\text{LLM}}$ generates a response $p$ for the query $q$ using the retrieved $K$ memories. Formally:
\begin{gather}
\{ \boldsymbol{m}_k \}_{k=1}^K = \pi_\theta^{\text{ret}}(q, \mathcal{M}, K), \quad
p= \pi_\theta^{\text{LLM}}\left(q, \{ \boldsymbol{m}_k \}_{k=1}^K \right).
\end{gather}

For efficiency, $\pi_\theta^{\text{ret}}$ and $\pi_\theta^{\text{LLM}}$ can be unified into a single policy that performs both memory retrieval and response generation for query $q$. 
\end{itemize}
% \item \textbf{Global System Objective:}

The goal of the memory system is to optimize downstream task performance. 
Given a dataset $\mathcal{D}$ of input pairs $(\mathcal{H}, q)$ and an evaluation function $\mathcal{F}$, we aim to learn an optimal policy set $\Pi^*$ that maximizes the expected performance:
\begin{gather}
\label{eq: objective_mmas}
\Pi^* = \arg\max_{\Pi} \mathbb{E}_{(\mathcal{H}, q) \sim \mathcal{D}} 
\left[ \mathcal{F} \big( p^*, \Pi(\mathcal{H}, q) \big) \right].
\end{gather}
Here, $\Pi = \{\pi_\theta^{\text{cons}_g}, \pi_\theta^{\text{ret}}\}$ denotes the policy set, and $p^*$ is the ground-truth response.

\subsection{Reinforcement Learning for LLMs}
\label{sec: rl_grpo}
To optimize LLMs' policy $\pi_{\theta}$, Reinforcement Learning (RL) \cite{RLHF,PPO} is typically used to maximize the expected reward while maintaining stability relative to a reference model $\pi_{\text{ref}}$. In this work, we primarily build upon Group Relative Policy Optimization (GRPO) \cite{GRPO}, an efficient RL algorithm that eliminates the need for a separate value function (critic) through group-based relative rewards. Given the input $x$, GRPO samples a group of outputs $\left\{o_1, \ldots, o_G\right\}$ from the current policy $\pi_{\theta}$ and optimizes the objective function:
\begin{equation}
\begin{aligned}
\mathcal{J}_{GRPO}\left(\pi_\theta\right)=\frac{1}{G} \sum_{i=1}^G \frac{1}{\left|o_i\right|} \sum_{t=1}^{\left|o_i\right|}\left\{\min \left[\rho_{i, t} \hat{A}_{i, t}, \hat{\rho}_{i, t} \hat{A}_{i, t}\right]-\beta \mathbb{D}_{KL}\right\},
\end{aligned}
\end{equation}
where $\rho_{i, t}=\frac{\pi_\theta\left(o_{i, t} \mid x, o_{i,<t}\right)}{\pi_{\theta_{\text {old }}}\left(o_{i, t} \mid x, o_{i,<t}\right)}$ is the importance sampling ratio, $\hat{\rho}_{i, t}=\operatorname{clip}\left(\rho_{i, t}; 1-\epsilon, 1+\epsilon\right)$ is the clipped ratio. The advantage $\hat{A}_{i, t}$ for each output $o_i$ is computed by normalizing its reward $r_i$ relative to the group mean and the standard deviation. Formally, we have: 
\begin{gather}
\hat{A}_{i, t} = \frac{r_i - \text{mean}(r)}{\text{std}(r)}, \text{ and }\quad 
\mathbb{D}_{KL} = \frac{\pi_{ref}(o_{i,t}|x, o_{i,<t})}{\pi_{\theta}(o_{i,t}|x, o_{i,<t})} - \log \frac{\pi_{ref}(o_{i,t}|x, o_{i,<t})}{\pi_{\theta}(o_{i,t}|x, o_{i,<t})} - 1.
\end{gather}

\section{Method}

\begin{figure*}[t]
  \centering
  \includegraphics[width=0.95\linewidth]{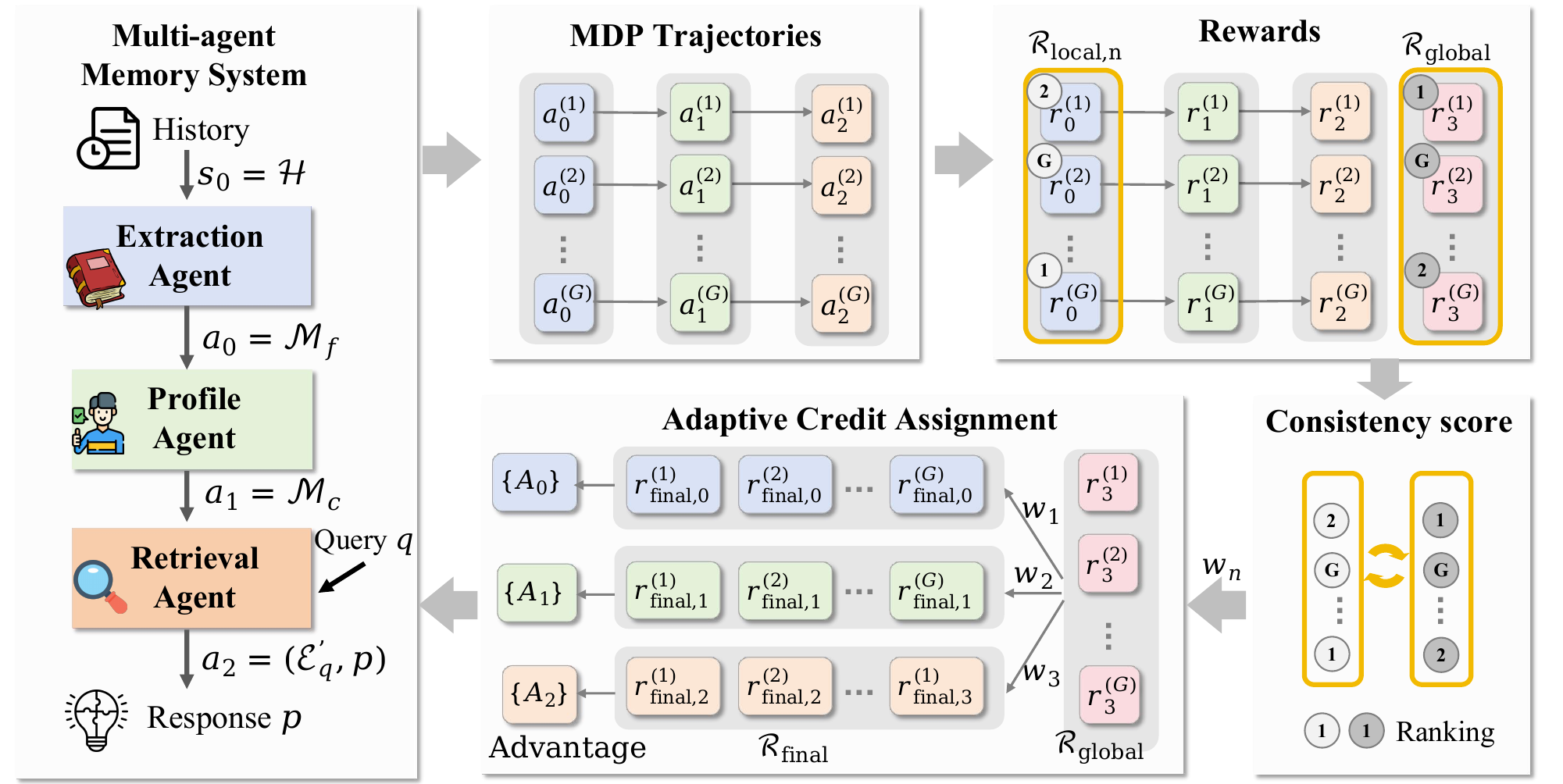}
  % \vspace{-4mm}
  \caption{The overview of our joint optimization framework for the multi-agent memory system, CoMAM, which promotes collaboration among agents via end-to-end reinforcement learning and an adaptive credit assignment mechanism.
}
  \label{fig:method}
  % \vspace{-6mm}
\end{figure*}
In this section, a joint optimization framework that enables agents to co-adapt via end-to-end reinforcement learning and an adaptive credit assignment mechanism, as shown in Figure \ref{fig:method}. We begin by defining the task-specific agents alongside the local and global rewards in Sec \ref{sec: definition of agents}. Then, we frame the heterogeneous multi-agent pipeline as a Markov decision process (MDP) in Sec \ref{sec: execution pipeline}, which models the inter-agent dependencies for joint training. Finally, we present the adaptive credit assignment mechanism in Sec \ref{sec: credit assignment}, which quantifies individual agent contributions and provides more targeted feedback for different agents.

\subsection{Task-Specific Agents Definition}
\label{sec: definition of agents}

Building on existing multi-agent memory paradigms \cite{hu2025memory,li2025cam}, we define a standard memory system consisting of three specialized agents: Extraction Agent, Profile Agent, and Retrieval Agent. 

\textbf{Extraction Agent} extracts fine-grained memory $\mathcal{M}_f$ with policy $\pi_\theta^{\text{cons}_f}$ from the raw conversational history $\mathcal{H}$. Formally, we define the set of evident information relevant to query $q$ as $\mathcal{E}_q \subseteq \mathcal{H}$. The agent’s objective is to maximize relevant evidence coverage with the reward function:
\begin{gather}
r_{\text{cons}_f} = \alpha \cdot \frac{|{\mathcal{M}_f} \cap \mathcal{E}_q|}{|\mathcal{E}_q|} + (1-\alpha) \cdot \frac{|{\mathcal{M}_f} \cap \mathcal{E}_q|}{|{\mathcal{M}_f} \cup \mathcal{E}_q|},
\label{eq: agent_extract}
\end{gather}
where $r_{\text{cons}_f}$ balances the coverage of relevant information and the filtering of irrelevant content. To prioritize fact coverage during the memory construction stage, $\alpha$ is set to $0.8$.

\textbf{Profile Agent} abstracts coarse-grained memory $\mathcal{M}_c$ (\eg user's preferences or behavioral patterns) from $\mathcal{M}_f$, guided by policy $\pi_\theta^{\text{cons}_c}$. The agent aims to maximize abstraction rationality, quantified by the model-based reward function:
\begin{gather}
r_{\text{cons}_c} = \mathcal{V}(\mathcal{M}_c, \mathcal{H}) \in [0,1],
\label{eq: agent_profile}
\end{gather}
where $r_{\text{cons}_c}$ evaluates the quality of $\mathcal{M}_c$ relative to the conversation history $\mathcal{H}$ according to predefined rubrics (\eg scenario, preference, and user profile). More details are presented in Appendix \ref{app: implementation}. $\mathcal{V}$ denotes a frozen pre-trained LLM for reward generation. Finally, the comprehensive memory set is defined as $\mathcal{M}=\{\mathcal{M}_f, \mathcal{M}_c\}$.

\textbf{Retrieval Agent} retrieves top-$k$ relevant memories $\mathcal{E}_q'=\{\boldsymbol{m}_k\}_{k=1}^K$ from memory set $\mathcal{M}$ and generate response $p$ for user's subsequent query $q$, guided by policy $\pi_\theta^{\text{ret}}$. To evaluate the retrieval performance, we define a rule-based reward function that enables precise matching between the recalled $\mathcal{E}_q'$ and ground-truth evidence $\mathcal{E}_q$:
\begin{gather}
r_{\text{ret}} = \beta \cdot \frac{|\mathcal{E}_q'\cap \mathcal{E}_q|}{|\mathcal{E}_q|} + (1-\beta) \cdot \frac{|\mathcal{E}_q'\cap \mathcal{E}_q|}{|\mathcal{E}_q'\cup \mathcal{E}_q|},
\end{gather}
where $r_{\text{ret}}$ balances the coverage and precision of retrieval. To prioritize retrieval precision while minimizing noise for answer generation, $\beta$ is set to $0.2$.

To evaluate the memory system's global performance, we define the rule-based reward for $\pi_\theta^{\text{ret}}$ to evaluate its answering accuracy for query $q$. Formally, let $p^*$ denote the ground-truth response for the query $q$, we have: 
\begin{gather}
r_{\text{ans}}=\mathbb{I}\left(p^*=p\right)
\end{gather}
where $\mathbb{I}$ denotes the indicator function that measures alignment between the generated response $p$ and ground-truth answer $p^*$, yielding $1$ if $p^*=p$ and $0$ otherwise.

Notably, the number and tasks of these agents can be further expanded to meet diverse application requirements in a multi-agent memory system.

\subsection{Inter-agent Dependency Modeling}
\label{sec: execution pipeline}

In a typical memory system, construction agents first convert conversational histories into structured memories, which are then consumed by retrieval agents to answer queries. This naturally forms a sequential pipeline for heterogeneous agents to coordinate through their intermediate outputs \cite{RL_MDP}. 
To capture such inter-agent dependencies during co-adaptation, we model the pipeline as a sequential Markov decision process (MDP), enabling joint optimization across heterogeneous agents. 
This formulation serves as a \emph{training-time abstraction}: it does not assume synchronous deployment, but provides a structured way to expose how upstream decisions influence downstream behavior.
Formally, the MDP is defined as $(\mathcal{S}, \mathcal{A}, T, R)$, where each step corresponds to a heterogeneous agent executing its policy $\{\pi_\theta^n\}_{n=0}^2$ (extraction, profile, retrieval). 
Given an input $(\mathcal{H}, q)$, the system evolves as a trajectory $\tau = (s_0, a_0, s_1, a_1, s_2, a_2)$:

\begin{itemize}[leftmargin=*]
\item $\mathcal{S}$: The state represents intermediate outputs of each agent in the MDP, with $s_0=\mathcal{H}$, $s_1=\mathcal{M}_f$, $s_2=(\mathcal{M}, q)$, and $s_3=(\mathcal{E}_q', p)$.

\item $\mathcal{A}$: Each action $a_n$ corresponds to the execution of agent $\pi_\theta^n$, including fine-grained memory extraction, coarse-grained abstraction, and retrieval with response generation.

\item $T$: The transition function $s_{n+1}=T(s_n, a_n)$ propagates intermediate outputs, where each agent’s output becomes part of the next agent’s input, explicitly exposing inter-agent dependencies.

\item $R$: Each step is associated with a local task reward $r_n$, together with a shared global reward $r_3$ reflecting downstream performance.
\end{itemize}

Within this formulation, CoMAM embeds inter-agent dependencies into state transitions along the optimization trajectory. The multi-agent memory system operates sequentially
to generate local and global rewards, enabling end-to-end
RL for heterogeneous policies' co-adaptation.

\subsection{Adaptive Credit Assignment Mechanism}
\label{sec: credit assignment}

While modeling inter-agent dependencies during joint training enables co-adaptation, relying solely on a shared global reward provides limited and ambiguous guidance for optimizing specialized agents \cite{hong2025multi,naive_assign}. 
In particular, when agents cannot distinguish their individual impact on the overall outcome, they receive indistinguishable learning signals, which prevent the development of effective role-specific policies and consequently hinder agents' collaboration in the memory system. 
To address this limitation, we introduce an adaptive credit assignment mechanism that distributes the global reward among agents in proportion to their estimated contribution, providing more informative training signals for each specialized agent.

Formally, for each input $(\mathcal{H}, q)$, we sample a group of MDP trajectories $\{\tau_i\}_{i=1}^G$, and collect local rewards for each agent $n$, $\mathcal{R}_{\text{local},n} = \{r_n^{(i)}\}_{i=1}^G$, along with the corresponding global rewards, $\mathcal{R}_{\text{global}} = \{r_3^{(i)}\}_{i=1}^G$, where $n \in \{0,1,2\}$. 
To estimate each agent's contribution to the global system, we measure the alignment between its local and global rewards using a ranking-based consistency metric, instantiated as Normalized Discounted Cumulative Gain (NDCG) \cite{NDCG_1}:
\begin{gather}
v_n = \text{NDCG}(\sigma(\mathcal{R}_{\text{local},n}), \sigma(\mathcal{R}_{\text{global}})),
\end{gather}
where $\sigma(\cdot)$ denotes the ranking operator. 
Intuitively, this metric captures whether trajectories with higher global rewards are also associated with higher local rewards for a given agent. 
Agents whose local reward rankings are more consistent with global outcome rankings receive higher credit, as their behaviors are more aligned with improving overall performance. 
Moreover, compared to value-based metrics, ranking-based measures are more robust under sparse and noisy rewards, as they rely on relative ordering rather than absolute magnitudes \cite{ranking_reward}.
The resulting scores are normalized to obtain adaptive weights $w_n$ with $\sum_n w_n = 1$: $
w_n = \frac{\exp(v_n)}{\sum_{n'} \exp(v_{n'})}$.
Each agent's final optimization signal combines its local reward with its adaptively weighted share of the global reward:
\begin{gather}
r_{\text{final},n}^{(i)} = r_n^{(i)} + w_n \cdot r_3^{(i)}.
\end{gather}

We then optimize each agent simultaneously using the GRPO algorithm (as introduced in Sec. \ref{sec: rl_grpo}) on these integrated rewards: $
\mathcal{J}_{\text{GRPO}}\left(\pi_\theta^n\right) = 
\frac{1}{G} \sum_{i=1}^G \min\Big[\rho_n^{(i)} A_n^{(i)}, \; \text{clip}(\rho_n^{(i)}, 1\pm\epsilon) A_n^{(i)}\Big]$,
where $\rho_n^{(i)}$ is the importance sampling ratio and $A_n^{(i)}$ the normalized advantage computed from $r_{\text{final},n}^{(i)}$. 
By leveraging both local task-specific rewards and adaptively weighted global rewards, this mechanism fosters collaboration while preserving agent specialization.

\section{Experiments}
In this section, we conduct extensive experiments to validate the effectiveness of our proposed CoMAM by addressing the following questions:
\begin{itemize}[leftmargin=*, itemsep=0pt, parsep=0pt, topsep=0pt]
\item RQ1: How does CoMAM perform against leading memory systems?
\item RQ2: What is the contribution of each agent within the memory system?
\item RQ3: What is the impact of inter-agent dependency modeling for joint optimization?
\item RQ4: What is the contribution of adaptive credit assignment for the specialization agents?
\item RQ5: What is the computational cost of the CoMAM?

% \item Can joint optimization boost the performance of local task specialization in the memory system?
\end{itemize}

\begin{table*}[t]
\centering
\caption{Query-Answering Accuracy on the PersonaMem Benchmark across 32K, 128K, and 1M context lengths. \textbf{Bold} indicates the best performance, and \underline{underlining} denotes the second best.  }
% \vspace{2pt}
\label{table: main_results}
\setlength{\tabcolsep}{14pt}
\begin{small}
\resizebox{\textwidth}{!}{%
\begin{tabular}{l|ccc|ccc}
\toprule
\multirow{2}{*}{Method} & \multicolumn{3}{c|}{Qwen} & \multicolumn{3}{c}{Llama} \\
\cmidrule{2-7}
& 32K & 128K & 1M & 32K & 128K & 32M \\
\midrule
\rowcolor{gray!15}& \multicolumn{6}{c}{\textbf{No memory}}\\
Base&0.41&0.39&0.38&0.35&0.31&0.32\\
RAG \cite{RAG} & 0.48 & 0.45 & 0.41 & 0.43 & 0.39 & 0.36 \\
\midrule
\rowcolor{gray!15} &\multicolumn{6}{c}{\textbf{Prompt-based memory}}\\
CAM \cite{li2025cam} & 0.53 & 0.50 & 0.45 & 0.48 & 0.45 & 0.43\\
MemoryBank \cite{memorybank} & 0.51 & 0.47 & 0.44 & 0.45 & 0.42 & 0.39 \\
A-Mem \cite{xu2025amem} & 0.50 & 0.44 & 0.45 & 0.46 & 0.37 & 0.36 \\
\midrule
\rowcolor{gray!15} &\multicolumn{6}{c}{\textbf{RL-based memory}}\\
Mem1 \cite{mem1} & \underline{0.59} & 0.57 & 0.58 & 0.56 & 0.58 & \underline{0.61} \\
Memory-R1 \cite{mem-r1} & 0.58 & \underline{0.60} & \underline{0.60} & \underline{0.57} & \underline{0.61} & 0.60 \\
\midrule
Ours & \textbf{0.64} & \textbf{0.70} & \textbf{0.66} & \textbf{0.62} & \textbf{0.68} & \textbf{0.69} \\
\rowcolor{gray!15} Impro&+8.5\%&+16.7\%&+10\%&+8.8\%&+11.5\%&+13.1\%\\
\bottomrule
\end{tabular}% 
}
\end{small}
\label{tab:main_results}
\end{table*}

\begin{wraptable}{t}{0.4\textwidth} 
\vspace{-10pt}
\centering
\small
\setlength{\tabcolsep}{8pt}
\caption{Results on LongMemEval using Qwen2.5-7B.}
\begin{tabular}{lccc}
\toprule
\textbf{Method} & $F_1 \uparrow$ & $B_1 \uparrow$ & $J_1 \uparrow$ \\
\midrule
RAG   & 18.27 & 14.57 & 22.20 \\
SeCom & 19.03 & 16.57 & 10.00 \\
A-Mem & 41.55 & 36.58 & 54.80 \\
Mem0  & 38.44 & 34.53 & 46.80 \\
GAM   & 51.75 & 48.96 & 52.00 \\
\cmidrule(lr){1-4}
Ours& 62.17 & 51.88 & 57.60\\
\bottomrule
\end{tabular}
\label{tab:longmemeval}
\end{wraptable}

\subsection{Experimental Settings}
\label{sec: exp_settings}
\paragraph{Dataset}
We conduct experiments on the PersonaMem \cite{personamem} and LongMemEval \cite{longmemeval} benchmarks. 
PersonaMem provides long-horizon conversational histories with varying context lengths (\ie 32K, 128K, and 1M tokens), each paired with seven downstream query types. 
LongMemEval focuses on multi-session conversational settings, requiring tracking information and performing temporal reasoning across sessions. They offer a more realistic evaluation of long-horizon memory construction and retrieval. 
Dataset processing details are provided in Appendix~\ref{app: dataset}.

\paragraph{Baselines}
We conduct a comprehensive evaluation of CoMAM against multiple leading baselines, including methods without a memory system ``No memory'', methods of building memory systems via dedicated prompts ``Prompt-based memory'', and methods of optimizing the memory systems with RL ``RL-based memory''. Detailed information of baselines are provided in Appendix \ref{app: exp_seetings}.
\begin{itemize}[leftmargin=*, itemsep=1pt, parsep=1pt, topsep=0pt]
\item \textbf{No memory:} ``Base'' represents directly injecting raw conversations as LLMs' input for query answering; RAG \cite{RAG} answers queries by retrieving raw conversational history;
\item \textbf{Prompt-based memory:} A-Mem \cite{xu2025amem} instantiates a memory system with construction, evolution, and retrieval agents; CAM \cite{li2025cam} and MemoryBank \cite{memorybank} build multi-level (fine-to-coarse granularity) memory via multi-agent dedicated prompts.
\item \textbf{RL-based memory:} Mem1 \cite{mem1} uses RL to optimize agents for memory searching during reasoning; Memory-R1 \cite{mem-r1} separately optimizes memory construction and retriever agents via RL.
\end{itemize}

\paragraph{Implementation Details.}
To evaluate CoMAM’s effectiveness and generalizability, we conduct experiments on PersonaMem and LongMemEval using LLM backbones from the Qwen \cite{qwen} and Llama \cite{Llama} families. Specifically, the extraction agent, profile agent, and reward model $\mathcal{V}$ are instantiated with Qwen2.5-3B-Instruct and Llama-3.2-3B-Instruct, while the retrieval agent leverages Qwen2.5-7B-Instruct and Llama-3.1-8B-Instruct, with model sizes tailored to each agent’s task complexity. All LLM backbones are configured with a 32K-token context window. Following existing work \cite{personamem,mem-r1,mem1}, we evaluate memory systems using query-answer accuracy. 
Specifically, we adopt multiple-choice accuracy for PersonaMem, and $F_1$, $B_1$, and $J_1$ (LLM-as-a-Judge) for LongMemEval. Table \ref{tab:main_results} presents the main results for both Qwen and Llama; other tables show only Qwen for simplicity. All tabulated results are averages of at least three independent runs with distinct random seeds.
More details are provided in the Appendix \ref{app: implementation}.

\subsection{Overall Performance (RQ1)}
To validate CoMAM’s efficacy for the memory system, we compare its query-answer accuracy against leading memory system baselines on the PersonaMem benchmark across its three history length settings (\ie 32K, 128K, and 1M tokens). As shown in Table \ref{table: main_results}, CoMAM delivers superior performance across all context length configurations (32K, 128K, and 1M) and two LLM backbone families (Qwen and Llama), fully confirming CoMAM's effectiveness and generalizability in long-horizon conversational scenarios.
Notably, RL-based memory baselines outperform prompt-based memory systems, validating that reinforcement learning is effective for optimizing memory system agents. For instance, the top prompt-based baseline CAM \cite{li2025cam} achieves only $0.45$-$0.53$ accuracy across contexts, while the RL-based Memory-R1 \cite{mem-r1} maintains a higher accuracy range of $0.57$–$0.61$. Moreover, CoMAM outperforms the state-of-the-art baseline Memory-R1, which employs independent agent optimization. 
This highlights the critical importance of joint optimization for multi-agent memory systems to achieve collaboration. Figure \ref{fig: seven_type} in Appendix \ref{app: question types} details Qwen-based results for the seven question types of PersonaMem. We further evaluate CoMAM on the LongMemEval benchmark. As shown in Table \ref{tab:longmemeval}, our method achieves the best performance across all metrics, consistently outperforming strong baselines such as GAM \cite{GAM}, further demonstrating its effectiveness in long-context memory retrieval and reasoning scenarios.

\begin{table*}[t]
\centering

\begin{minipage}[t]{0.47\textwidth}
\centering
\setlength{\tabcolsep}{12pt}
\renewcommand{\arraystretch}{0.85}
\caption{Ablation study for the contribution of different agents to the memory system.}
\label{table: abla_agents}
\begin{small}
\begin{tabular}{l|ccc}
\toprule
\multirow{2}{*}{Method} & \multicolumn{3}{c}{Accuracy (\%)} \\
\cmidrule{2-4}
& 32K & 128K & 1M \\
\midrule
Base & 0.23 & 0.26 & 0.37 \\
w/o Ex & 0.52 & 0.52 & 0.51 \\
w/o Pr & 0.57 & 0.52 & 0.57 \\
w/o Re & 0.54 & 0.51 & 0.51 \\
\midrule
no train & 0.50 & 0.46 & 0.44 \\
w/o train Ex & 0.57 & 0.61 & 0.54 \\
w/o train Pr & 0.57 & 0.61 & 0.64 \\
w/o train Re & 0.62 & 0.66 & 0.59 \\
\midrule
Ours & \textbf{0.64} & \textbf{0.70} & \textbf{0.66} \\
\bottomrule
\end{tabular}
\end{small}
\end{minipage}
\hspace{0.03\textwidth}
\begin{minipage}[t]{0.47\textwidth}
\centering
\caption{Ablation study of joint optimization for heterogeneous agents. }
\label{tab: abla_MDP}
\begin{small}
\begin{tabular}{l|ccc}
\toprule
\multirow{2}{*}{Method} & \multicolumn{3}{c}{Accuracy (\%)} \\
\cmidrule{2-4}
& 32K & 128K & 1M \\
\midrule
\rowcolor{gray!15} & \multicolumn{3}{c}{\textbf{Single policy}} \\
no train & 0.51 & 0.48 & 0.45 \\
+ independent RL & 0.57 & 0.56 & 0.54 \\
+ end-to-end RL & 0.60 & 0.59 & 0.59 \\
\midrule
\rowcolor{gray!15} & \multicolumn{3}{c}{\textbf{Heterogeneous policy}} \\
no train & 0.50 & 0.46 & 0.44 \\
+ independent RL & 0.57 & 0.61 & 0.59 \\
+ end-to-end RL (ours) & \textbf{0.64} & \textbf{0.70} & \textbf{0.66} \\
\bottomrule
\end{tabular}
\end{small}
\end{minipage}
\end{table*}

\subsection{Ablation for Agents (RQ2)}
\begin{figure}[t]
\centering
\vspace{-2mm}
  \includegraphics[width=0.78\columnwidth]{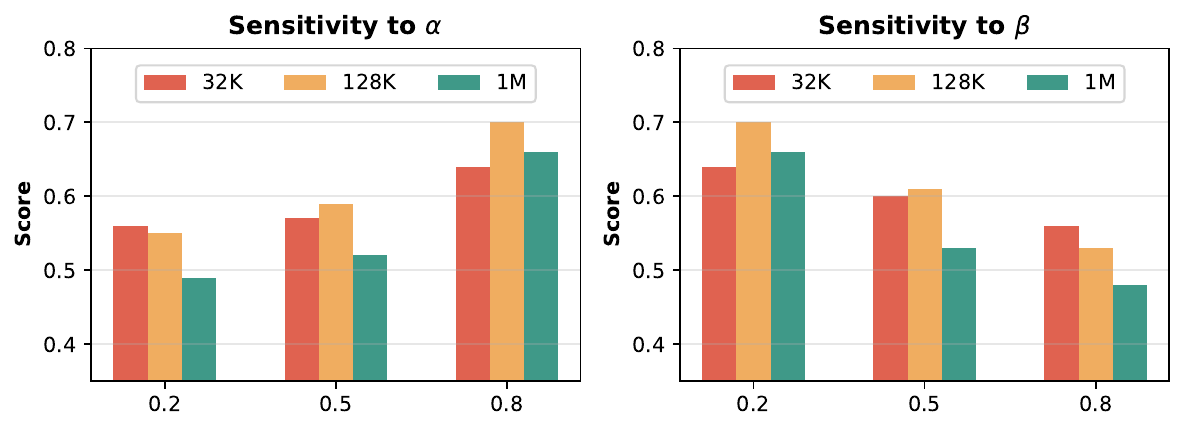}
  \vspace{-2mm}
  \caption{Sensitivity of CoMAM to the hyperparameters $\alpha$ and $\beta$ as introduced in Sec. \ref{sec: definition of agents}.
  % where $\alpha$ controls the trade-off between information coverage and noise filtering for the extraction agent, and $\beta$ balances recall and precision for the retrieval agent.
}
  \label{fig: sen_alpha_beta}
\vspace{-6mm}
\end{figure}

To verify the necessity and contribution of the three core agents (Extraction Agent, Profile Agent, and Retrieval Agent), we designed targeted ablation experiments, with results presented in Table \ref{table: abla_agents}.
First, the ``Base'' baseline (without memory agents) achieves the lowest performance across all context lengths, confirming the significance of memory systems for long-horizon conversations. Second, removing any agent can lead to significant performance degradation, whether the Extraction Agent (``w/o Ex'') that builds factual fine-grained memories, the Profile Agent (``w/o Pr'') that captures core user preferences, or the Retrieval Agent (``w/o Re'') that searches for memories to answer queries. These results demonstrate that all three agents play complementary and unique roles in supporting the memory system.
Furthermore, the ``w/o train'' variants (full agent architecture with one agent left untrained) provide additional validation: untraining any single agent results in 2–8\% performance drops compared to CoMAM, highlighting that optimizing these agents is critical to building an effective memory system. We further analyze the sensitivity of the reward coefficients $\alpha$ and $\beta$ in Figure \ref{fig: sen_alpha_beta}, showing that the chosen settings ($\alpha=0.8$, $\beta=0.2$) enable the extraction agent to better preserve key information and the retrieval agent to effectively filter noise.

\subsection{Ablation for Dependency Modeling (RQ3)}
To model inter-agent dependencies for joint optimization, we formulate the heterogeneous multi-agent pipeline as a sequential MDP, which enables end-to-end RL. To validate its efficacy, we design ``single policy'' (all agents share one policy across tasks \cite{memagent,share_one_policy}) and ``heterogeneous policy'' (each agent has a dedicated policy per task) variants, with two training strategies: ``independent RL'' (fine-tunes one agent’s policy at a time, others fixed without updates) and ``end-to-end RL'' (simultaneously updates policies via MDP trajectory for joint optimization). As shown in Table \ref{tab: abla_MDP}, ``single policy'' variants consistently underperform ``heterogeneous policy'' variants, confirming that a shared policy undermines agent specialization by integrating diverse task-specific objectives. Furthermore, ``+end-to-end RL'' outperforms ``independent RL'' across both policy paradigms. This suggests that sequential MDP modeling during end-to-end RL training acts as a structured abstraction to capture how upstream decisions influence downstream behavior, thereby enabling more effective co-adaptation among agents.

\subsection{Ablation for Adaptive Credit Assignment (RQ4)}
\begin{figure}[t]
\centering
\vspace{-2mm}
  \includegraphics[width=0.78\columnwidth]{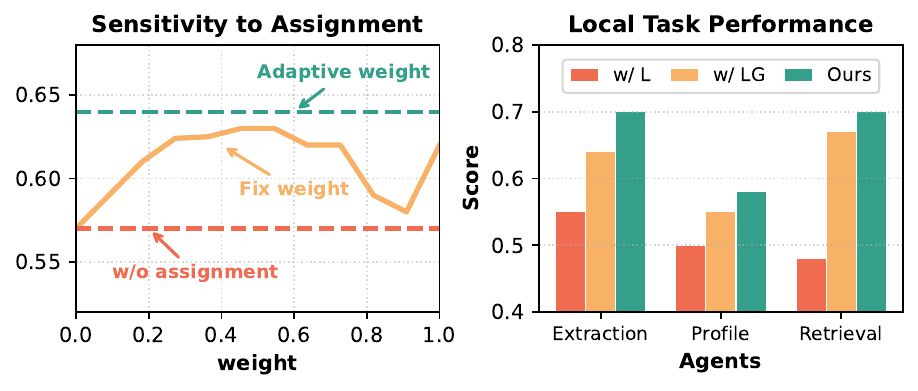}
  \vspace{-2mm}
  \caption{The left demonstrates CoMAM’s sensitivity to the credit assignment weight, and the right shows the credit assignment’s impact on each local agent’s performance on the 32K setting.
}
  \label{fig: abla_credit}

\end{figure}

To address RQ4, we evaluate different credit assignment strategies to assess the effectiveness of CoMAM’s adaptive mechanism. As shown in Table~\ref{table: abla_adaptive}, optimizing agents using only local rewards (``end-to-end RL w/ L'') or only global rewards (``end-to-end RL w/ G'') improves over the ``no train'' and ``independent RL'' variants, but underperforms approaches that combine both signals (``end-to-end RL w/ LG'' and ``Ours''). This suggests that integrating local and global rewards is essential for balancing agent specialization and collaboration.
Notably, ``Ours'', with adaptive credit assignment, consistently outperforms ``end-to-end RL w/ LG'', which uses uniform global rewards, highlighting the advantage of assigning credit based on agents' contributions. As further shown in Figure \ref{fig: abla_credit}, adaptive weights $w_n$ outperform fixed weighting schemes.
Moreover, adaptive credit assignment improves task-specific performance: our method consistently achieves higher local task performance across all agents compared to both ``w/ LG'' and ``w/ L'', indicating stronger role-specific capability. Additional experimental results and analyses of the reward curves are provided in Appendix \ref{app: exp_reward}.

\subsection{Efficiency (RQ5)}
We further evaluate the training efficiency of CoMAM by comparing the convergence steps of joint optimization and independent optimization across 32K, 128K, and 1M context length settings, with results presented in Table~\ref{tab: training_cost}. The total convergence steps serve as the core metric to reflect training efficiency, where fewer steps indicate higher efficiency in reaching stable performance.
Specifically, independent optimization trains each agent separately, with total steps being the sum of individual convergence steps for the Extraction, Profile, and Retrieval Agents. In contrast, joint optimization trains all agents in parallel, with total steps determined by the slowest-converging agent, resulting in fewer total steps. While individual agents under joint training may take more steps for convergence (\eg 58 vs.\ 48 steps for the Extraction Agent at the 32K setting), joint training shortens the total training duration.
Importantly, per-step cost is similar in both settings, since frozen agents in independent optimization still participate in the forward pass to complete the pipeline and compute rewards. Thus, the efficiency gain mainly comes from reducing redundant forward passes, as agents update simultaneously in joint optimization.
Moreover, inference latency is decoupled from the training paradigm: agents optimized by joint or independent optimization achieve comparable inference efficiency at deployment. Yet joint optimization delivers superior global performance via inter-agent collaboration and adaptive credit assignment, highlighting the distinct advantage of joint multi-agent optimization for memory systems.

\begin{table*}[t]
\centering

\begin{minipage}[t]{0.465\textwidth}
\centering
\caption{Ablation study for different credit assignment strategies.}
\label{table: abla_adaptive}
\begin{small}
\setlength{\tabcolsep}{7pt}
\begin{tabular}{l|ccc}
\toprule
\multirow{2}{*}{Method} & \multicolumn{3}{c}{Accuracy (\%)} \\
\cmidrule{2-4}
& 32K & 128K & 1M \\
\midrule
no train & 0.50 & 0.46 & 0.44 \\
+ independent RL & 0.57 & 0.61 & 0.59 \\
\midrule
end-to-end RL w/ L & 0.57 & 0.65 & 0.61 \\
end-to-end RL w/ G & 0.60 & 0.64 & 0.60 \\
end-to-end RL w/ LG & 0.62 & 0.65 & 0.63 \\
\midrule
Ours & \textbf{0.64} & \textbf{0.70} & \textbf{0.66} \\
\bottomrule
\end{tabular}
\end{small}
\end{minipage}
\hfill
\begin{minipage}[t]{0.505\textwidth}
\centering
\setlength{\tabcolsep}{6pt}
\renewcommand{\arraystretch}{0.82}
\caption{Convergence step comparison of the joint and independent optimization.}
\label{tab: training_cost}
\begin{small}
\setlength{\tabcolsep}{5pt}
\begin{tabular}{l|cccc}
\toprule
Method & Extract & Profie & Retrieval & Total \\
\midrule
\rowcolor{gray!15} & \multicolumn{4}{c}{\textbf{32K}} \\
Independent & 48 & 57 & 65 & 160 \\
Joint & 58 & 32 & 68 & 68 \\
\midrule
\rowcolor{gray!15} & \multicolumn{4}{c}{\textbf{128K}} \\
Independent & 210 & 280 & 300 & 790 \\
Joint & 300 & 350 & 320 & 350 \\
\midrule
\rowcolor{gray!15} & \multicolumn{4}{c}{\textbf{1M}} \\
Independent & 285 & 180 & 150 & 415 \\
Joint & 210 & 300 & 310 & 310 \\
\bottomrule
\end{tabular}
\end{small}
\end{minipage}

\end{table*}

\section{Conclusion}
In this work, we study the limitations of independently optimizing agents in multi-agent memory systems, highlighting the lack of co-adaptation and the ambiguity of credit assignment under a shared global objective. To address these challenges, we propose CoMAM, a joint optimization framework that enables agents' co-adaptation via end-to-end reinforcement learning, together with an adaptive credit assignment mechanism for more informative training signals. By modeling the inter-agent dependencies as an MDP and integrating local and global rewards adaptively, CoMAM facilitates collaboration across agents. Experimental results demonstrate consistent improvements over strong baselines, validating the effectiveness of joint optimization with adaptive credit assignment for multi-agent memory systems. The limitation is discussed in Appendix \ref{app: limitation}. 

% \newpage
% \section*{Impact Statements}
% This paper presents work whose goal is to advance the field of machine learning. There are many potential societal consequences of our work, none of which we feel must be specifically highlighted here.

\clearpage
\bibliographystyle {plain}
\bibliography{reference}

\newpage
\newpage
\appendix

\section{Algorithm}

Here we present the algorithm of CoMAM’s training process in Algorithm \ref{alg:comam}.
\begin{algorithm}
\caption{Collaborative Reinforcement Learning Framework  for Multi-Agent Memory Systems (CoMAM)}
\begin{algorithmic}[1]
\label{alg:comam}
\REQUIRE $\mathcal{H}$, $q$, $\mathcal{E}_q$, $p^*$, $\{\pi_\theta^{\text{cons}_f}, \pi_\theta^{\text{cons}_c}, \pi_\theta^{\text{ret}}\}$, $\alpha=0.8$, $\beta=0.2$, $G$, $\epsilon$
\ENSURE Optimized $\{\pi_\theta^{\text{cons}_f}, \pi_\theta^{\text{cons}_c}, \pi_\theta^{\text{ret}}\}$, response $p$

% ========== Phase 1: Multi-Agent Execution (Single Trajectory) ==========

\FOR{$i = 1$ \textbf{to} $G$}
   \STATE $\mathcal{M}_f \leftarrow \pi_\theta^{\text{cons}_f}(\mathcal{H})$ \hfill $\triangleright$ Extract fine-grained memory
\STATE $r_{\text{cons}_f} \leftarrow \alpha \cdot \frac{|\mathcal{M}_f \cap \mathcal{E}_q|}{|\mathcal{E}_q|} + (1-\alpha) \cdot \frac{|\mathcal{M}_f \cap \mathcal{E}_q|}{|\mathcal{M}_f \cup \mathcal{E}_q|}$
\STATE $\mathcal{M}_c \leftarrow \pi_\theta^{\text{cons}_c}(\mathcal{M}_f)$ \hfill $\triangleright$ Abstract coarse-grained memory
\STATE $r_{\text{cons}_c} \leftarrow \mathcal{V}(\mathcal{M}_c, \mathcal{H})$
\STATE $\mathcal{M} \leftarrow \{\mathcal{M}_f, \mathcal{M}_c\}$
\STATE $(\mathcal{E}_q', p) \leftarrow \pi_\theta^{\text{ret}}(\mathcal{M}, q)$ \hfill $\triangleright$ Retrieve and generate response
\STATE $r_{\text{ret}} \leftarrow \beta \cdot \frac{|\mathcal{E}_q' \cap \mathcal{E}_q|}{|\mathcal{E}_q|} + (1-\beta) \cdot \frac{|\mathcal{E}_q' \cap \mathcal{E}_q|}{|\mathcal{E}_q' \cup \mathcal{E}_q|}$
\STATE $r_{\text{ans}} \leftarrow \mathbb{I}(p^* = p)$ \hfill $\triangleright$ Global reward (Downstream task accuracy)
\ENDFOR
\STATE Collect local reward $r_n^{(i)}$ for agent $n$, add to $\mathcal{R}_{\text{local},n}$
\STATE Collect global reward $r_{\text{ans}}^{(i)}$, add to $\mathcal{R}_{\text{global}}$
\STATE $v_n \leftarrow \text{NDCG}(\sigma(\mathcal{R}_{\text{local},n}), \sigma(\mathcal{R}_{\text{global}}))$ \hfill $\triangleright$ Consistency score
\STATE $w_n \leftarrow \frac{\exp(v_n)}{\sum_{n'} \exp(v_{n'})}$ \hfill $\triangleright$ Adaptive credit weights
\STATE $r_{\text{final},n}^{(i)} \leftarrow r_{\text{ans}}^{(i)} + w_n \cdot r_n^{(i)}$ \hfill $\triangleright$ Final reward for agent $n$

% ========== Phase 3: Joint RL Optimization (GRPO) ==========
\FOR{$n = 0$ \textbf{to} $2$} 
    \FOR{$i = 1$ \textbf{to} $G$} 
        \STATE $A_n^{(i)} \leftarrow \text{Normalize}(r_{\text{final},n}^{(i)})$ \hfill $\triangleright$ Compute advantage
        \STATE $\rho_n^{(i)} \leftarrow \frac{\pi_\theta^n(a_n^{(i)} \mid s_n^{(i)})}{\pi_{\theta_{\text{old}}}^n(a_n^{(i)} \mid s_n^{(i)})}$ \hfill $\triangleright$ Importance sampling ratio
    \ENDFOR

    \STATE $\pi_\theta^n \leftarrow \text{Update}(\pi_\theta^n, \mathcal{J}_{\text{GRPO}}(\pi_\theta^n))$ \hfill $\triangleright$ Optimize agent policy
\ENDFOR

\end{algorithmic}
\end{algorithm}

\section{Details of Experimental Settings}
\label{app: exp_seetings}
\subsection{Dataset}
\label{app: dataset}

In the PersonaMem benchmark, each interaction history consists of 10, 20, or 60 multi-turn sessions (corresponding to 32k, 128k, or 1M tokens) to form varying context lengths. Each session contains 15–30 conversation turns, designed to naturally reveal or update user preferences. In addition, each history is paired with a suite of subsequent queries across seven types, enabling comprehensive evaluation of the memory system. The question types include: (1) Recall user-shared facts, (2) Suggest new ideas, (3) Acknowledge latest user preferences, (4) Track full preference evolution, (5) Revisit reasons for preference updates, (6) Provide preference-aligned recommendations, and (7) Generalize to new scenarios. Compared to existing benchmarks \cite{LOCOMO,longmemeval}, PersonaMem’s strengths in long context history and different query type enable rigorous assessment of memory systems in conversational scenarios.
For PersonaMem’s three context length settings (32k, 128k, and 1M tokens), we partition their training and validation sets based on these queries. The statistical breakdown of training and validation queries is summarized in Table \ref{table:dataset}. 

In the LongMemEval benchmark \cite{longmemeval}, each sample consists of a scalable multi-session user–assistant interaction history, where information is gradually revealed and may be revised across sessions. The benchmark contains 500 manually created questions and defines two standard settings: LongMemEval-S with about 115K tokens per sample, and LongMemEval-M with 500 sessions and roughly 1.5M tokens. Its evaluation targets five core memory abilities—information extraction, multi-session reasoning, temporal reasoning, knowledge updates, and abstention—and the released question categories include single-session-user, single-session-preference, single-session-assistant, multi-session, temporal-reasoning, and knowledge-update. Compared with benchmarks that mainly test long-context recall, LongMemEval more explicitly evaluates cross-session reasoning and dynamic user-state tracking in realistic assistant settings.

To process the training set, we first identify the specific session within the full conversational history that contains explicit evidence relevant to each query $q$, and we define this session as the query’s targeted conversational history $\mathcal{H}$. We then construct $(\mathcal{H},q)$ data pairs required for generating individual MDP trajectories as the core training data. The position of each target session within the complete interaction history is specified in the original benchmark. Ground-truth evidence $\mathcal{E}_q$ for each query $q$ is extracted using the Seed-1.6 model and validated through human evaluation. We emphasize that $\mathcal{E}_q$ is only used for computing training rewards and is not available during inference, ensuring no information leakage. While obtaining $\mathcal{E}_q$ involves additional annotation effort, this cost is incurred only during training and does not affect inference-time efficiency, making it a practical and acceptable design choice. For evaluation, multiple distinct queries may share the same full conversational history that includes multiple sessions. This design enables further evaluation in complex inference scenarios where memory construction agents operate on a full conversational history with multiple sessions, while retrieval agents are activated for subsequent queries at varying frequencies.

\subsection{Detailed Baselines}

Here, we detailed our baseline methods, including methods without memory, prompt-based memory systems, and RL-optimized memory systems.

\textbf{Methods without memory:}
\begin{itemize}[leftmargin=*]
\item Base denotes the direct injection of raw conversational data as input to large language models (LLMs) for query answering tasks. Specifically, when the length of the input conversation exceeds the context window constraint, the input is truncated at the context window limit.
\item RAG \cite{RAG} answers queries by retrieving raw conversational history via embedding-based similarity matching. The embedding model employed is all-MiniLM-L6-v2, and ultra-long input contexts are uniformly partitioned into fixed-length segments, each consisting of 2,048 tokens.
\end{itemize}
\textbf{Prompt-based memory systems}
\begin{itemize}[leftmargin=*]
\item A-Mem \cite{xu2025amem} instantiates a memory system through multiple agents with dedicated prompts, including the memory retrieval construction, precursor memory evolution, and retrieval agents; 
\item CAM \cite{li2025cam} builds a hierarchical memory system via multiple agents, where dedicated agents undertake node construction (via ego-centric disentanglement and node replication) and information aggregation (for hierarchical abstraction). We implement it on the PersonaMem benchmark.
\item MemoryBank \cite{memorybank} construct a human-like memory for LLMs that uses a forgetting curve to remember and forget information over time. Similarly, we implement it on the PersonaMem benchmark.
\item SeCom \cite{secom} designs a modular memory system for long-horizon conversational agents, where dedicated components handle memory construction and retrieval using prompt-based multi-agent.
\item Mem0 \cite{Mem0} instantiates a multi-agent memory system that maintains fine-grained and coarse-grained user memories, performing extraction and retrieval with specialized prompts. 

\end{itemize}
\textbf{RL-based memory:} 
\begin{itemize}[leftmargin=*]
\item GAM \cite{GAM} introduces an agentic memory system with a memorizer for key information and a researcher for retrieval and integration at runtime, trained with reinforcement learning.
\item Mem1 \cite{mem1} proposes a reinforcement learning (RL) framework that empowers agents to conduct efficient multi-turn information retrieval and reasoning for query answering. The optimized agents prioritize direct memory retrieval from history, rather than memory construction.
\item Memory-R1 \cite{mem-r1} constructs a Memory Manager that learns optimal memory operations (ADD, UPDATE, DELETE) and an Answer Agent tasked with distilling relevant information from retrieved entries. Uniquely, the system is optimized via independent reinforcement learning (RL) in two stages: the Memory Manager is optimized first, followed by the Answer Agent.
\end{itemize}

\subsection{More Implementation Details}
\label{app: implementation}
For a fair comparison, all the baselines are implemented with the same LLM backbones as our CoMAM, including Qwen2.5-7B-Instruct and Llama-3.1-8B-Instruct. All experiments are conducted on a cluster of 8 NVIDIA A100 GPUs. For the implementation of variant ``+independent RL'', each agent is trained by fusing its task-specific rewards and global accuracy rewards, with the remaining agents frozen. For the ``single policy'' variant, we adopt a shared policy to serve as the policy for all three agents during training and inference. The prompts for specializing different agents and rubrics for evaluating the profile agent are presented below. The temperature of LLMs is set to $1$ for training and $0.8$ for evaluation. 
The detailed hyperparameter settings for CoMAM are presented in Table \ref{tab: hyper}. 

\begin{table}[t]
\vspace{-2mm}
\renewcommand\arraystretch{1.1}
\caption{The statistics of queries after the train-validation split across different dataset settings.}
% \vspace{2mm}
\label{table:dataset}
\centering
\setlength{\tabcolsep}{2.8mm}
\small
\begin{tabular}{cccc}
\toprule
Dataset& 32K& 128K& 1M
 \\
 \midrule
\#Train &313& 1,744& 1,830\\
\#Validate& 100 &200& 200\\
\#question types& 7& 7 &7\\
% \cline{2-3}
\toprule
% \vspace{-6mm}
\end{tabular}
\end{table}

\begin{table}[t]
\vspace{-2mm}
\renewcommand\arraystretch{1.1}
\caption{Hyperparameters of CoMAM on PersonaMem benchmark across different context lengths.}
% \vspace{2mm}
\label{tab: hyper}
\centering
\setlength{\tabcolsep}{2.8mm}
\small
\begin{tabular}{cccc}
\toprule
Dataset& 32K& 128K& 1M
 \\
 \midrule
batch\_size &128&128&128\\
group\_size& 8 &8& 8\\
max\_model\_length& 32k& 32k &32k\\
epochs&5&5&5\\
max\_output\_tokens&3k&3k&3k\\
\toprule
% \vspace{-6mm}
\end{tabular}
\end{table}

\begin{tcolorbox}[colback=black!5!white, colframe=black!75!white, title=Fine-grained memory construction -- Extraction Agent]
   
You are the Memory Extractor. Your task is to analyze the upcoming Conversation History. 

Retain important conversations as much as possible in order to answer future questions about the conversation history, removing only the redundant text of each sentence.

The conversation history is provided below:

Conversation History: 
\begin{verbatim}
   Historical Conversations
\end{verbatim}
Output *only* the retained conversations from the conversation history you process.
    
\end{tcolorbox}

\vspace{5pt}
\begin{tcolorbox}[colback=black!5!white, colframe=black!75!white, title=Coarse-grained memory construction -- Profile Agent]

You are the Profile Abstractor. You will receive fine-grained memory information from the Memory Extractor.
Your task is to analyze the memory you receive and abstract it into higher-level knowledge for the user profile (such as scenarios, user behavior patterns, and preferences).
Fine-grained Memory:
\begin{verbatim}
   Fine-grained Memory
\end{verbatim}
Output only the abstracted user profile from the memory you receive.
    
\end{tcolorbox}

\vspace{5pt}
\begin{tcolorbox}[colback=black!5!white, colframe=black!75!white, title=Memory retrieval and query answer -- Retrieval Agent ]
You are the Retrieval agent. You have received fine-grained memory from the Memory Extractor and coarse-grained memory from the Profile Abstractor. 

Your task is to retrieve relevant memory to answer the following question accurately.

Multi-grained Memory:

\begin{verbatim}
   Fine-grained Memory from the Memory Extractor
   Coarse-grained memory from the Profile Abstractor
\end{verbatim}

Question:
\begin{verbatim}
   User's Subsequent Query
\end{verbatim}

Options:
\begin{verbatim}
   The Four Options
\end{verbatim}
     
Please output the retrieved relevant information in <information> </information> and choose the most appropriate option (a), (b), (c), or (d) in <final\_answer></final\_answer>.
    
\end{tcolorbox}

\vspace{5pt}
\begin{tcolorbox}[colback=black!5!white, colframe=black!75!white, title=Rubrics for Profile Agent Reward Evaluation]

You are an evaluation expert. Your task is to assess whether the provided \textit{Coarse-grained memory} accurately captures the key information from the given \textit{Conversation History}.

Conversation History:
\begin{verbatim}
  User--LLM conversational history
\end{verbatim}

Coarse-grained memory:
\begin{verbatim}
  Output generated by the Profile Agent
\end{verbatim}

\textbf{Evaluation Criteria:}
\begin{enumerate}
    \item Are the scenarios sufficiently rich and comprehensive?
    \item Is the analysis of user preferences sufficiently in-depth?
    \item Is the user persona sufficiently detailed and representative?
\end{enumerate}

\textbf{Output Format:}

Please output a single scalar score within \texttt{<score></score>} as a float between 0.0 and 1.0, without any additional text.

\begin{itemize}
    \item \textbf{1.0}: The synthesized insight perfectly captures the scenario, user preferences, and persona.
    \item \textbf{0.0}: The synthesized insight fails to capture the relevant information.
\end{itemize}

\end{tcolorbox}
\newpage

\section{More Detailed Experimental Results}
\begin{figure*}[t]
\vspace{-2mm}
  \includegraphics[width=\columnwidth]{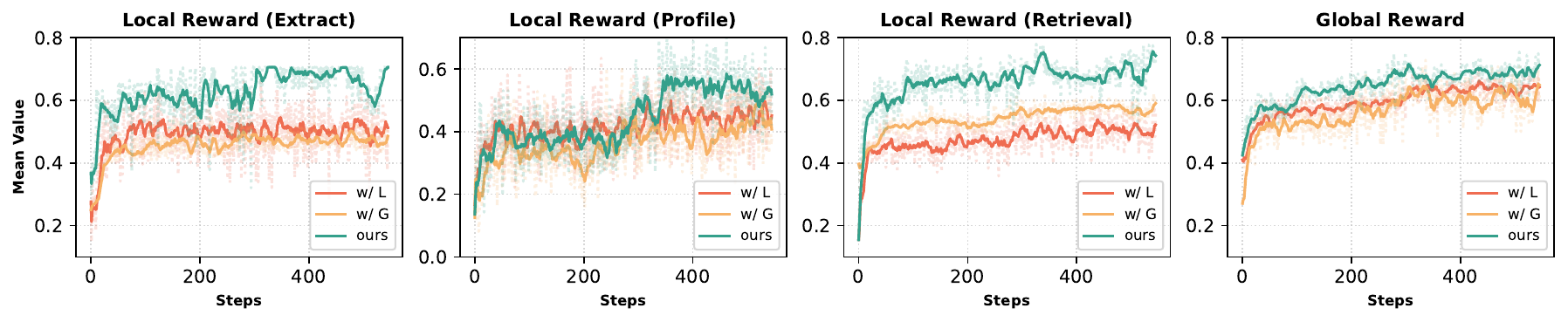}
  \vspace{-2mm}
  \caption{The curves for local task-specific reward $r_n$ and global accuracy reward $r_3$.
}
  \label{fig: reward_curve}
\vspace{-4mm}
\end{figure*}
\subsection{The Impact of Joint Optimization on Local Task Specialization}
\label{app: exp_reward}
Joint optimization with adaptive credit assignment not only enhances global system performance but also promotes the specialization of each agent’s local tasks. As illustrated in Figure \ref{fig: reward_curve}, we compare the local task reward curves of agents under three training strategies: optimizing with only local rewards (``w/ L''), optimizing with only global rewards (``w/ G''), and our joint optimization with adaptive credit assignment (``Ours'').
Key observations are as follows:
\begin{itemize}[leftmargin=*]
\item Superior local reward convergence: For all three agents (Extraction, Profile, Retrieval), the local task rewards of ``Ours'' consistently outperform ``w/ L'' and ``w/ G'' in the stable phase. This indicates that adaptive credit assignment will not compromise local task capabilities; instead, it strengthens them by aligning local improvement with global goals.

\item Global performance enhancement: The global reward curve of ``Ours'' converges to 0.70, significantly higher than ``w/ L' (0.65) and ``w/ G'' (0.64), proving that joint optimization realizes global performance enhancement by fostering inter-agent collaboration.
\end{itemize}

Adaptive credit assignment mechanism quantifies the correlation between each agent’s local performance and the global outcome, ensuring that agents receive rewards not only for completing local tasks well but also for contributing to the global goal. This guides agents to optimize their local strategies in a way that benefits the entire system, thus achieving both local specialization and inter-agent collaboration for the global memory system.

\subsection{Detailed Performance Across Question Types}

\label{app: question types}
\begin{figure}[H]
% \vspace{-2mm}
  \includegraphics[width=\columnwidth]{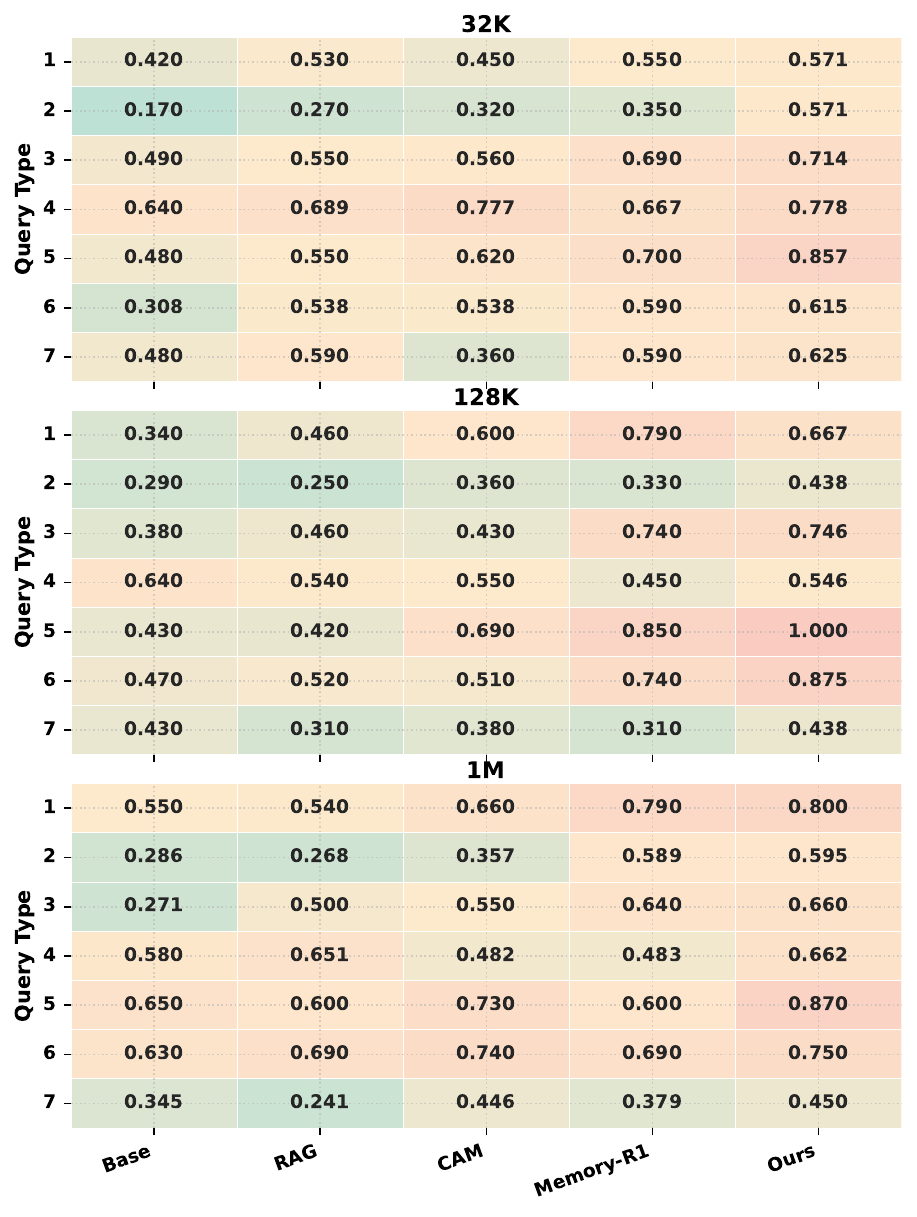}
  \vspace{-2mm}
  \caption{Detailed performance of different methods across seven question types on the PersonaMem benchmark across three history length settings (\ie 32K, 128K, and 1M). The number of question types is consistent with those listed in Section \ref{sec: exp_settings}
}
  \label{fig: seven_type}
\vspace{-4mm}
\end{figure}

Figure \ref{fig: seven_type} details Qwen-based results for the seven question types, and we observe that CoMAM performs well in both detailed factual recall (\eg Type 1) and abstract preference-based question types (\eg Type 6), providing additional validation of its efficacy.

\newpage

\newpage
\section{Discussion}
\subsection{Difference with existing memory framework}
Existing memory systems and our CoMAM differ fundamentally in optimization paradigm and reward design:
\begin{itemize}[leftmargin=*]
\item
\textbf{Optimization Paradigm:} Independent Optimization vs. Joint Optimization

Existing frameworks primarily adopt independent optimization. For example, MIRIX \cite{mirix} designs a meta-agent and six specialized agents for memory management, optimizing each via tailored prompts; Memagent \cite{memagent} focuses on a single retrieval agent, training it with query-answer accuracy as the reward; Memory-R1 \cite{mem-r1} splits the system into a Memory Manager and Answer Agent, training them in separate stages—first optimizing the Manager with the Agent frozen, then freezing the Manager to optimize the Agent. This approach ignores inter-agent independencies and lacks co-adaptation: memory construction is not adjusted based on together with retrieval behavior, and the agent can only adapt to a fixed Manager’s output, leading to insufficient inter-agent collaboration.

CoMAM proposes end-to-end joint optimization via modeling inter-agent dependencies in the MDP trajectory, where each agent’s output directly serves as the next agent’s input. All three agents are updated simultaneously during training, with inter-agent dependencies embedded into state transitions, ensuring they evolve collaboratively toward the global goal of the memory system.

\item \textbf{Reward Design:} Single Global Reward vs. Adaptive Credit Assignment

Existing frameworks that optimize memory agents with RL primarily rely solely on final query-answer accuracy as the reward for all agents. For example, Mem-$\alpha$ and Mem1 \cite{mem1} use this single reward to train construction and retrieval agents, respectively.
Memory-R1 \cite{mem-r1}, and GAM \cite{GAM} also adopt this design, using only answer accuracy to optimize their agents separately. This ambiguous credit assignment prevents agents from distinguishing their individual contributions, leading to unstable convergence.

CoMAM introduces adaptive integration of local and global rewards. We design task-specific local rewards (\eg Extraction Agent’s information coverage, Profile Agent’s abstraction rationality, Retrieval Agent’s retrieval precision) and combine them with the global reward (answer accuracy). Using NDCG to quantify the ranking consistency between each agent’s local rewards and the global reward, we dynamically allocate global credit based on this consistency. This ensures agents are rewarded both for excelling at their local tasks and contributing to global performance, avoiding ambiguous optimization.

\end{itemize}
\subsection{Difference with existing multi-agent system RL}

\begin{itemize}[leftmargin=*]
\item \textbf{Heterogeneous Policy vs. Homogeneous Policy}: Some multi-agent reinforcement learning (RL) methods adopt a shared policy across all agents. For instance, CORY \cite{uniform_merge} proposes an RL framework featuring ``Pioneer'' and ``Observer'' agents that \textbf{share a single policy model}. These agents co-evolve through multi-agent collaboration and periodic role-swapping. In contrast, considering the heterogeneity of agents in a memory system, CoMAM employs heterogeneous policies tailored to distinct agents. 

\item \textbf{Workflow: Sequential vs. Concurrent}: Existing multi-agent system (MAS) methods often adopt a debate-style pipeline for reasoning tasks, such as math tasks. For instance, MAPoRL \cite{MAPORL} models agent interactions as concurrent debate, where multiple agents generate responses in parallel, focusing on persuasive collaboration for general reasoning. In contrast, CoMAM explicitly models inter-agent dependencies as a sequential Markov Decision Process (MDP), providing a structured way to expose how upstream decisions influence downstream behavior. This ensures that memory construction quality dynamically adapts to retrieval requirements and vice versa.

\item \textbf{Collaboration Paradigm: Joint Optimization vs. Independent Alignment}: 
Existing methods, such as Optimas \cite{Optimas}, foster collaboration by aligning local rewards with global objectives via heuristic reward shaping for compound AI systems. However, it optimizes individual agents separately, leveraging locally-global aligned rewards rather than through end-to-end joint training. In contrast, CoMAM adopts a joint optimization paradigm that updates all agents simultaneously.

\item \textbf{Credit Assignment: Adaptive Assignment vs. Equal Allocation}: Most existing systems use static credit assignment mechanisms without quantifying agents’ varying contributions: AT-GRPO \cite{naive_assign} distributes rewards equally across agents using fixed weights, ignoring task differences; CoMAS \cite{xue2026comas} adopts a zero-sum adversarial reward scheme that pits agents against each other, which is misaligned with the cooperative nature of memory systems; In contrast, CoMAM assigns agent-specific credit based on a ranking-based proxy that measures the alignment between each agent’s local task rewards and the global system reward. This formulation captures whether trajectories with higher global performance are consistently associated with higher local rewards for a given agent. Agents with stronger alignment receive higher credit, as their behaviors are more predictive of overall system improvement.
This provides more informative training signals for specialized agents.
\end{itemize}

\section{Limitation}
\label{app: limitation}

While CoMAM demonstrates the effectiveness of joint optimization with adaptive credit assignment for multi-agent memory systems, several limitations remain.

\textbf{1. Simplified Agent Design.}  
We adopt a compact set of agents (Extraction, Profile, and Retrieval) to focus on core collaborative behaviors. More advanced functionalities, such as memory editing, consistency maintenance, and redundancy removal, are not explicitly modeled and remain for future exploration.

\textbf{2. Limited Cross-Agent Credit Modeling.}  
CoMAM estimates agent-specific credit by modeling the ranking consistency between each agent’s local rewards and the global system reward. However, relationships among agents’ local rewards are not explicitly considered in the credit estimation. Incorporating such cross-agent relationships may provide more informative credit signals and further improve coordination.

\section{Broader Impact}
CoMAM aims to improve the effectiveness of multi-agent memory systems for long-horizon conversational tasks by introducing a joint optimization framework with end-to-end reinforcement learning and adaptive credit assignment. This may benefit applications such as personalized assistants and long-context reasoning systems. 
However, it does not inherently address potential privacy risks associated with sensitive user information.
\onecolumn

% \newpage
% \input{checklist.tex}

\end{document}